\documentclass[preprint,showpacs,preprintnumbers,amsmath,amssymb]{revtex4}

\usepackage{graphicx}
\usepackage{dcolumn}
\usepackage{bm}

\begin{document}

\preprint{Opt. Commun. 254 (2005) 353-360}

\title{Amphoteric refraction at the interface between isotropic and anisotropic
media}

\author{Hailu Luo}\thanks{Author to whom correspondence should be addressed.
E-mail: hailuluo@163.com}
\author{Wei Hu}\thanks{Author to whom correspondence should be addressed.
E-mail: huwei@scnu.edu.cn}
\author{Xunong Yi}
\author{Haiying Liu}
\author{Jing Zhu}
\affiliation{Laboratory of Light Transmission Optics, South China
Normal University, Guangzhou 510630, China}
\date{\today}

\begin{abstract}
It is found that the amphoteric refraction, i.e. the refraction
can be either positive or negative depending on the incident
angles, could occur at a planar interface associated with a
uniaxially anisotropic positive index media (PIM) or an
anisotropic negative index media (NIM). Particularly, the
anomalous negative refraction can occur at a planar interface from
an isotropic PIM to an anisotropic PIM, whereas the anomalous
positive refraction occurs at the interface from an isotropic PIM
to an anisotropic NIM. The optimal conditions to yield the two
unusual refractions are obtained. The difference of the two types
of amphoteric refraction is discussed.
\end{abstract}

\pacs{78.20.Ci, 41.20.Jb, 42.25.Gy }
\keywords{negative index media, Anisotropic media, Amphoteric
refraction.}
\maketitle

\section{Introduction}
The material with negative index of refraction, also known as
left-handed media , first introduced by Veselago in 1968
\cite{Veselago1968}, has recently received much more attention in
literature. Experimentally, an artificial composite media,
meta-material, which has negative index of refraction in the
microwave regime, has been built and
measured\cite{Smith2000,Shelby2001}. Theoretically, the perfect
lens, which can realize sub-wavelength focusing, was first
proposed by Pendry \cite{Pendry2000}. Although some arguments have
been evoked on the concepts of negative refractive index and the
perfect lens
both\cite{Valanju2002,Hooft2001,William2001,Garcia2002}, the
phenomenon of negative refraction has been examined and confirmed
by many researchers
\cite{Parazzoli2003,Smith2002,Pacheco2002,Foteinopoulou2003,Houck2003,Santos2003,Feise2002}.

Negative refraction, in which the tangential component of the
Poynting vector changes sign when light refracted, is one of the
dramatic properties of the left-handed media. It occurs at the
interface between a normal positive index media (PIM), where the
electric permittivity $\epsilon$ and magnetic permeability $\mu$
are both positive, and a negative index media (NIM), where
$\epsilon$ and $\mu$ are both negative\cite{Veselago1968}.  It was
also found recently that negative refraction can occur in some
other media. Negative refraction in photonic crystal without
negative effective index of refraction has been
reported\cite{Notomi2000,Luo2002}. Lindell {\it et al}
\cite{Lindel2001} have shown that negative refraction can occur at
an interface associated with a uniaxially anisotropic media, where
only one of the four parameters of $\epsilon$ and $\mu$ tensors
are negative. Recently, Zhang {\it et al }\cite{Zhang2003} have
demonstrated that an amphoteric refraction, i.e. the refraction
can be either positive or negative depending on the incident
angles, may prevail at the interface between two anisotropic PIM
when their optical axes are in different directions. Further, Liu
{\it et al}\cite{Liu2004} have pointed out that the amphoteric
refraction, as well as the omnidirectional total transmission, can
occur at a planar interfaces associated with a uniaxial media.

Here we investigate the amphoteric refraction characteristics of
extraordinary light at the planar interface associated with an
anisotropic positive index media or an anisotropic negative index
media. It is found that the anomalous negative refraction can
occur at a planar interface from an isotropic PIM to an
anisotropic PIM, while the anomalous positive refraction occurs at
the interface from an isotropic PIM to an anisotropic NIM. Relying
on dispersion equation, the optimal optical axis angle and the
strongest anomalous refraction angle are obtained. It is noted
that for a strong anisotropy media, the maximum incident angle
yields the anomalous refraction could be very large. The
difference of amphoteric refraction associated with anisotropic
PIM and with anisotropic NIM is also discussed.

\section{Amphoteric refraction of light}
The negative refraction at an interface associated with uniaxially
anisotropic media, where some of the four parameters of $\epsilon$
and $\mu$ tensors are negative, has been generally discussed
\cite{Lindel2001,Hu2002,Smith2003}. However, in those literature,
the optical axis of uniaxial NIM was assumed to be normal or
parallel to the interface. Here we assume that there is an angle
$\varphi$  between the optical axis and the planar interface. It
is the necessary condition for the amphoteric refraction.

We consider the propagation of a planar wave of frequency $\omega$
as ${\bf E}={\bf E}_0 e^{i {\bf k} \cdot {\bf r}-i \omega t}$ and
${\bf H} = {\bf H}_0 e^{i {\bf k} \cdot {\bf r}-i\omega t}$,
through an isotropic PIM toward an uniaxially anisotropic media.
For the nonmagnetic isotropic PIM, the relative permittivity and
permeability are both scalars, i.e.  $\epsilon_I$ and $\mu_I$. For
the convenience and without loss of generality, we assume that the
permeability of uniaxially anisotropic media is isotropic, whereas
the relative permittivity is a second-rank tensor, given in
principal coordinates as
\begin{eqnarray}
\epsilon=\left(
\begin{array}{ccc}
\epsilon_\perp  &0 &0 \\
0 & \epsilon_\perp &0\\
0 &0 & \epsilon_\parallel
\end{array}
\right).\label{matrix}
\end{eqnarray}
We assume that  $\epsilon_\perp,\epsilon_\parallel>0$ and $\mu>0$
for the anisotropic PIM, and $\epsilon_\perp,\epsilon_\parallel<0$
and $\mu<0$ for the anisotropic NIM, in which both ordinary and
extraordinary indices are negative. As shown in Fig.1, we also
assume that all the wave vectors and the optical axis are in the
$x-z$ planar, where the interface is the $x-y$ plane at $z=0$. The
isotropic media is on the left hand side of interface, while the
anisotropic media on the right hand side. We refer to the angle of
the optical axis, $\varphi$, as the angle between the optical axis
of anisotropic media and the $x$ direction.

\begin{figure}
\includegraphics[width=8cm]{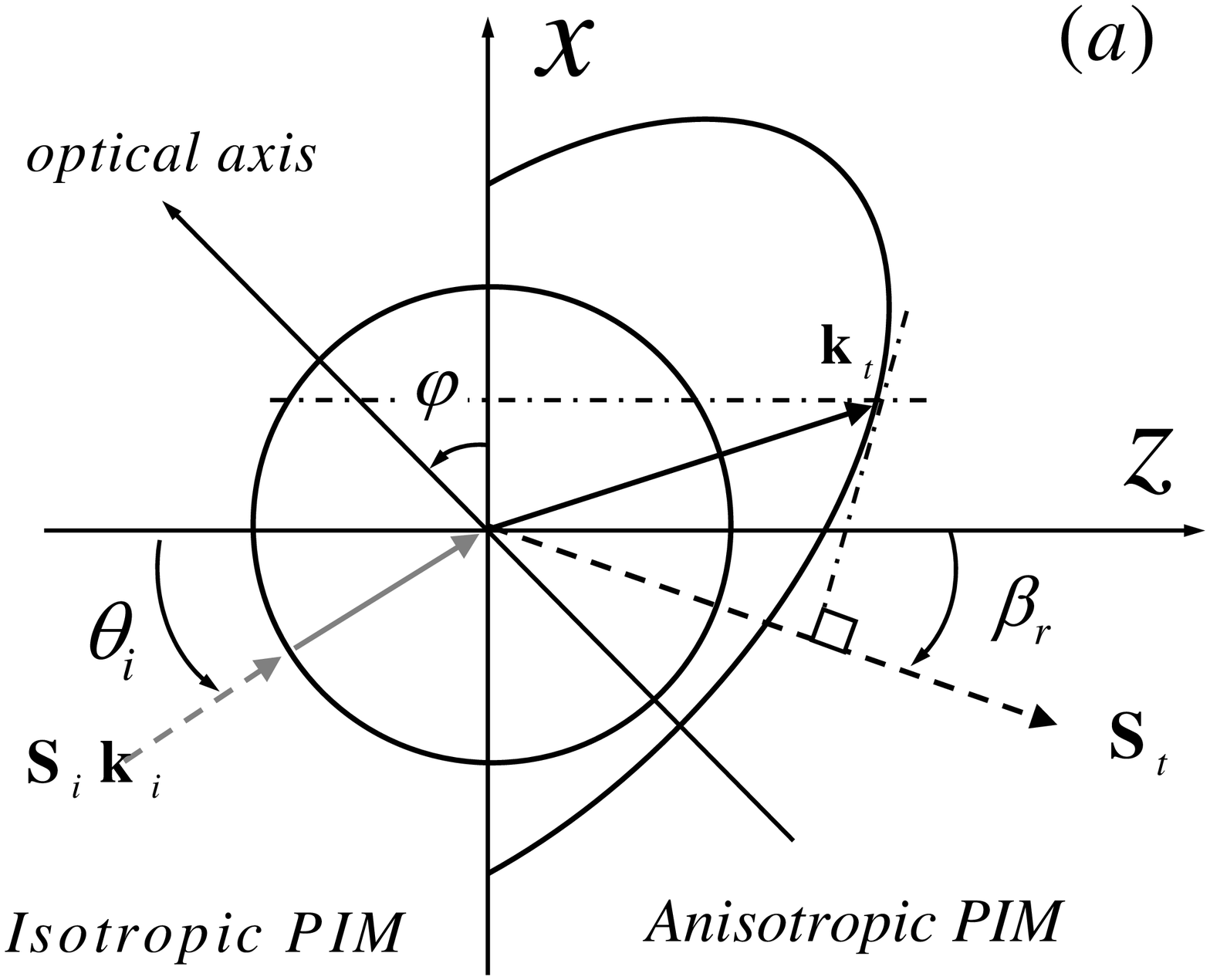}
\includegraphics[width=8cm]{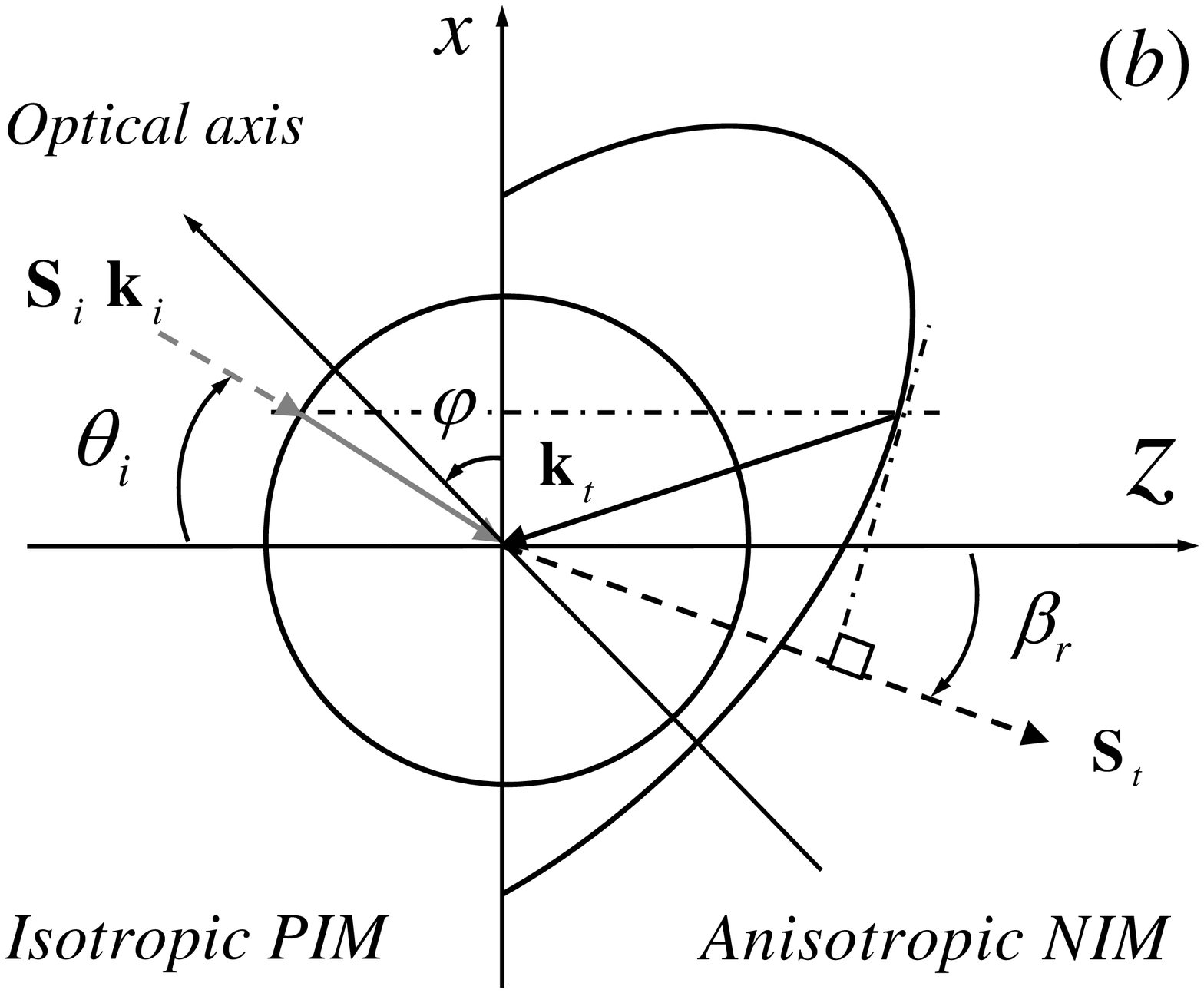}
\caption{\label{Fig1}The mechanisms for the amphoteric
refractions. (a) The negative refraction from an isotropic PIM to
an anisotropic PIM. (b) The positive refraction from an isotropic
PIM to an anisotropic NIM. The iso-frequency wave-vector surfaces
for the isotropic media (left) and the anisotropic media (right)
are circle and ellipse, respectively. The incident and transmitted
wave vectors ${\bf k}_i$ and ${\bf k}_t$ are indicated by the
solid arrow, whereas the energy flows ${\bf S}_i$ and ${\bf S}_t$
are indicated by dashed arrows. The $k_x$ components should be
continuous across the interface, and the vector ${\bf S}_t$ is
normal to the tangential surface of the iso-frequency wave-vector
ellipse at the point ${\bf k}_t $.}
\end{figure}

When dealing with an extraordinary plane wave, polarized along x
direction, one have $E_y = H_x = H_z = 0$. The iso-frequency
surface of the wave vector ${\bf k}_t$ for transmitted wave is an
ellipse described by the dispersion relation. It is worth pointing
out that for an extraordinary wave the time-averaged Poynting
vector for the transmitted wave ${\bf S}_t$, which defines the
actual direction of the energy flow of the extraordinary light, is
normal to the tangential surface of the wave vector ellipse at
point ${\bf k}_t $.

The dispersion relation for extraordinary wave in uniaxial media
can be easily obtained by solving Maxwell's equations for plane
waves in the principal coordinates\cite{Yariv1984} with a rotation
of coordinates on $y$-axis, as
\begin{equation}
\alpha k_{tx}^2+\beta k_{tz}^2+\gamma
k_{tx}k_{tz}=\frac{\omega^2}{c^2}, \label{dispersion}
\end{equation}
where $k_{tx}$ and $k_{tz}$ represent the $x$ and $z$ components
of transmitted wave vector ${\bf k}_t$ and $\alpha$, $\beta$ and
$\gamma$ are given by
\begin{eqnarray}
\alpha &=&\frac{1}{\epsilon_\perp \epsilon_\parallel\mu}
(\epsilon_\perp \sin^2\varphi+\epsilon_\parallel \cos^2\varphi)\nonumber\\
\beta &=&\frac{1}{\epsilon_\perp \epsilon_\parallel\mu}
(\epsilon_\perp \cos^2\varphi+\epsilon_\parallel \sin^2\varphi)\nonumber\\
\gamma &=&\frac{1}{\epsilon_\perp\epsilon_\parallel\mu}
(\epsilon_\perp \sin 2\varphi -\epsilon_\parallel \sin 2\varphi
).\nonumber
\end{eqnarray}
The $z$-compnent of the wave vector can be found by the solution
of Eq.(\ref{dispersion}), i.e.
\begin{eqnarray}
k_{tz}=\frac{1}{2\beta }\left[\sigma\sqrt{(\gamma -4\alpha\beta)
k_{tx}^2+4\beta \frac{\omega^2}{c^2} }-\gamma^2 k_{tx}
\right].\label{ktz}
\end{eqnarray}
Here $\sigma=1$ for  PIM  and $\sigma=-1$ for  NIM . This choice
of sign ensures that light power propagates away from the surface
to the $+z$ direction.

We now determine the angle of refraction. The incident angle of
light in the isotropic PIM is $\theta_i
=\tan^{-1}(k_{ix}/k_{iz})$, and the refractive angle of the
transmitted wave vector in anisotropic media can be found by
$\theta_r= \tan^{-1}(k_{tx}/k_{tz})$. From the boundary conditions
at the interface $z=0$, the tangential components of the wave
vectors must be continuous, i.e. $k_{tx}=k_{ix}$. Based on
Eq.(\ref{ktz}) and the boundary conditions, one can obtain
\begin{eqnarray}
\theta_r=\tan^{-1} \left[ \frac{2\beta \sqrt{\epsilon_I} \sin
\theta_i}{ \gamma \sqrt{\epsilon_I} \sin
\theta_i-\sigma\sqrt{(\gamma^2 -4\alpha\beta) \epsilon_I \sin^2
\theta_i +4\beta} }\right]\label{xktz}
\end{eqnarray}

It is well know that the actual propagation direction of an
extraordinary light in an anisotropic media differs from the
direction of its wave vector. In the following of this paper, the
direction of an extraordinary light is defined based on the
time-averaged Poynting vector ${\bf S}=\frac{1}{2} {\bf Re}({\bf
E}^\ast\times \bf{H})$, named as ray refractive angle. This ray
refractive angle is given by
\begin{eqnarray}
\beta_r=\tan^{-1}\left(\frac{  S_{tx} }{  S_{tz}}\right).
\label{energy}
\end{eqnarray}
For monochromatic extraordinary wave, it can be simplified as
$\beta_r=\tan^{-1} (-E_{tz}/E_{tx}) $. The $x$ and $z$ components
of the electric field ${\bf E}_t$ for transmitted wave can be
derived from the equation $\nabla \cdot{\bf D}=0 $  associated
with a rotation of coordinates, i.e.
\begin{eqnarray}
\epsilon_\perp(k_{tx}\cos\varphi-k_{tz}\sin\varphi)
(E_{tx}\cos\varphi-E_{tz}\sin\varphi) \nonumber\\
+ \epsilon_\parallel(k_{tx}\sin\varphi+k_{tz}\cos\varphi)
(E_{tx}\sin\varphi+E_{tz}\cos\varphi)=0. \label{ddd}
\end{eqnarray}
Based on Eq.(\ref{ktz}), (\ref{energy}), (\ref{ddd}), and the
boundary continuous conditions, one can derive the equation for
the ray refractive angle of extraordinary light,
\begin{eqnarray}
\beta_r=\tan^{-1} \left[ \frac{( 4\alpha\beta-\gamma^2)
\sqrt{\epsilon_I} \sin \theta_i+\sigma\gamma\sqrt{(\gamma^2
-4\alpha\beta) \epsilon_I \sin^2 \theta_i +4\beta} }{2\sigma\beta
\sqrt{(\gamma^2 -4\alpha\beta)\epsilon_I \sin^2 \theta_i
+4\beta}}\right].\label{br}
\end{eqnarray}

Due to the uniaxial anisotropy, the time-averaged Poynting vector
${\bf S}_t$ and the wave vector ${\bf k}_t$ are in different
directions. There is a bending angle between them. Because of the
bending angle, for some incident wave the negative refraction can
occurs in anisotropic PIM, while the positive refraction can
occurs in anisotropic NIM.

At the interface, there are two conditions that the light
refraction has to obey. The first is the continuous conditions of
the tangential components of the wave vectors, i.e.
$k_{tx}=k_{ix}$. The second is the energy conservation condition
which requests the normal components $S_{tz}$ and $S_{iz}$ must
have the same sign. For the anisotropic PIM shown in Fig.1a, the
normal components $S_{tz}$ and $k_{tz}$ are always in the same
sign, so we know that $k_{tz}$ and $k_{iz}$ have the same sign
too. It yields that the wave vector refraction is always positive.
In this case, if the tangential components $S_{tx}$ and $k_{tx}$
are in the opposite signs due to the bending angle, it leads to a
manifestation of the negative refraction. It is noted that the
negative refraction only occurs in a narrow range of incident
angles.

It is contrary in the anisotropic NIM as shown in Fig.1b. The
normal components $S_{tz}$ and $k_{tz}$ are in the opposite signs
now, then $k_{tz}$ and $k_{iz}$ have the different signs. So the
wave vector refraction is always negative. The positive refraction
also occurs when  $S_{tx}$ and $k_{tx}$ have the same sign due to
the bending angle. Therefore, the amphoteric refractions are
determined by whether the signs of $S_{tx}$ and $k_{tx}$ are same
or opposite. Their physical essential is due to the uniaxial
anisotropic. For the isotropic media, the Poynting vector ${\bf
S}_t$  is always parallel or anti-parallel with the wave vector
${\bf k}_t$ in PIM or NIM, respectively. The signs of tangential
components $S_{tx}$ and $k_{tx}$ are always same (positive
refraction for PIM) or opposite (negative refraction for NIM).
There is no amphoteric refraction for isotropic media.

\section{The realization of anomalous refractions}

As an example of amphoteric refraction in anisotropic PIM, the
numerical results calculated for a $YVO_4$ crystal, which has
recently been proposed by Zhang {\it et al} in their
experiment\cite{Zhang2003}, is shown in Fig.2a. The $YVO_4$
crystal is a nonmagnetic uniaxial crystal with
$\epsilon_\perp=n_0^2=4.07103$, $\epsilon_\parallel=n_e^2=5.06614$
and $\mu=1$. For comparison, an anisotropic NIM is shown in
Fig.2b, with parameters of $\epsilon_\perp=-4.07103$,
$\epsilon_\parallel=-5.06614$ and $\mu=-1$.

\begin{figure}
\includegraphics[width=8cm]{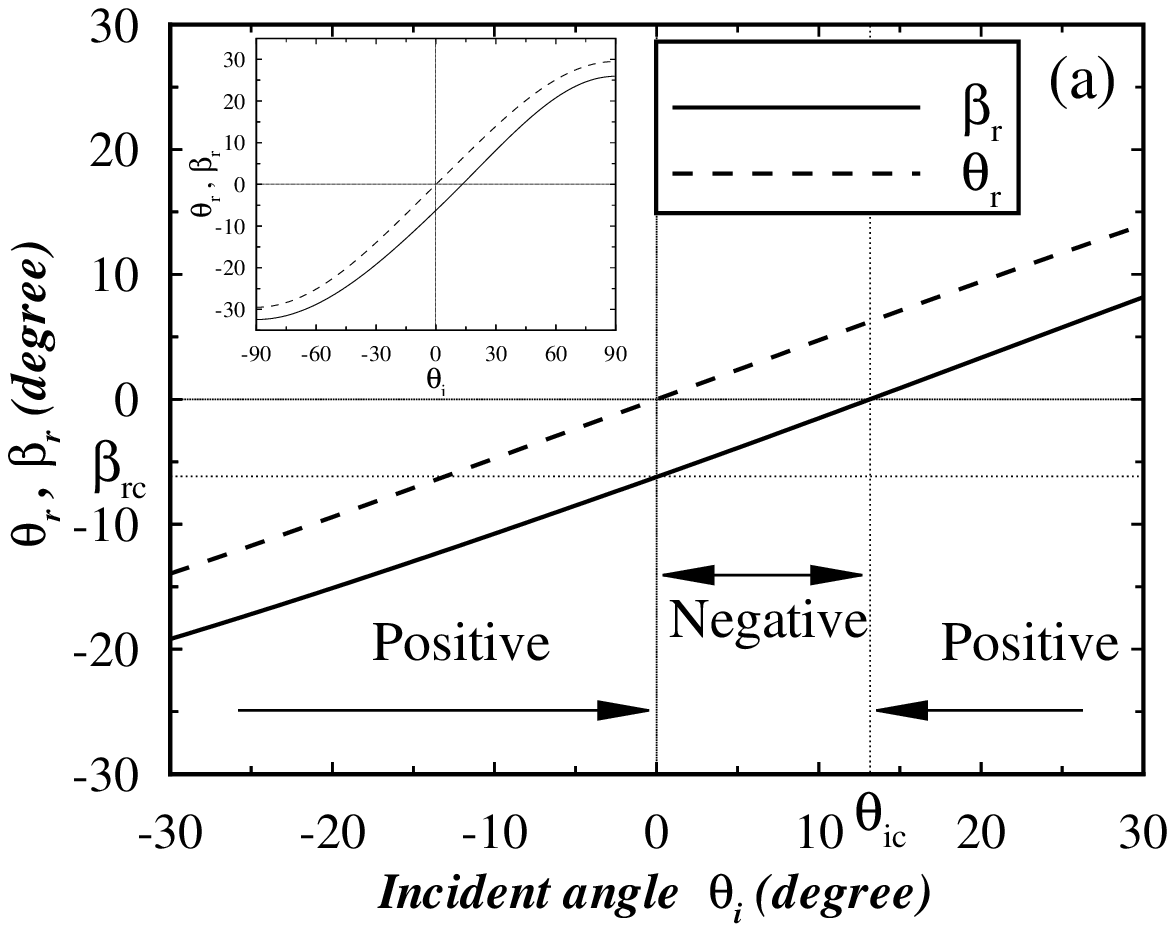}
\includegraphics[width=8cm]{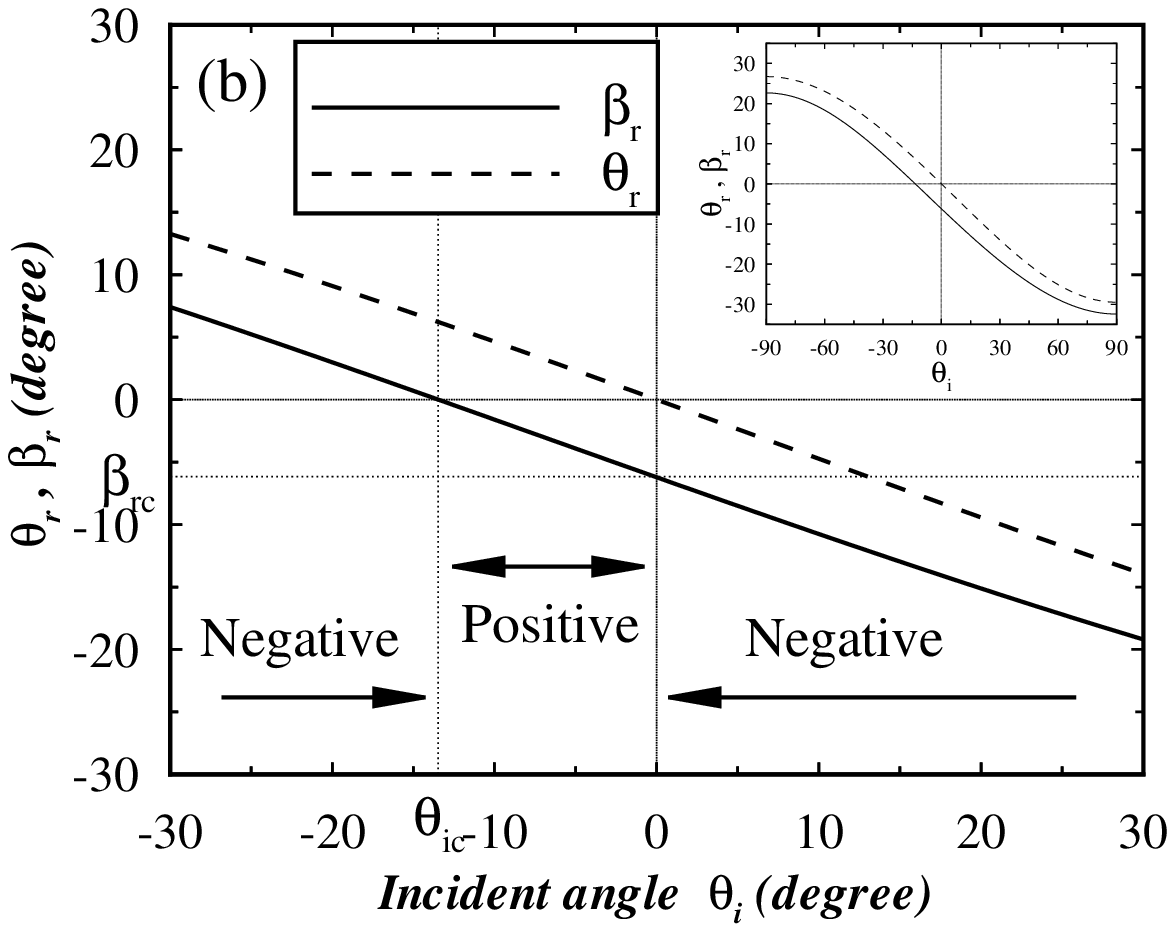}
\caption{\label{Fig2}The variations of the ray refractive angle
$\beta_r$ for energy flow and the transmitted wave vector angle
$\theta_r$ as a function of the incident angle $\theta_i$, where
$\varphi=\pi/4$. (a) In the anisotropic PIM, the anomalous
negative refraction ($\theta_i/\beta_r < 0$) occurs in region
$0\leq \theta_i \leq \theta_{ic}$. (b) In the anisotropic NIM, the
anomalous  positive refraction ($\theta_i/\beta_r > 0$) occurs in
region $0 \geq \theta_i \geq \theta_{ic}$. The inset figures are
same as main frames. }
\end{figure}

In Fig.2, the ray refractive angle $\beta_r$ and the wave-vector
refractive angle $\theta_r$ are shown as the solid and dashed
curves, respectively. The negative refraction occurs when curves
fall in the second or fourth quadrant of the $\theta_i-\beta_r$
plane. One can see in Fig.2a that it is always positive refraction
for the wave-vector refractive angle $\theta_r$. Whereas for the
ray refraction angle $\beta_r$, the incident angle can be divided
into three regions: one negative refraction ($\theta_i/\beta_r <
0$) and two positive refraction($\theta_i/\beta_r > 0$). The
negative refraction occurs in the fourth quadrant of the
$\theta_i-\beta_r$ plane, i.e. $0 \leq \theta_i \leq \theta_{ic}$.
From Fig.2b, it is contrary for anisotropic NIM. It is always
negative refraction for the wave-vector refractive angle
$\theta_r$ whereas positive refraction occurs in the third
quadrant for $\theta_{ic} \leq \theta_i \leq 0$.

\begin{figure}
\includegraphics[width=8cm]{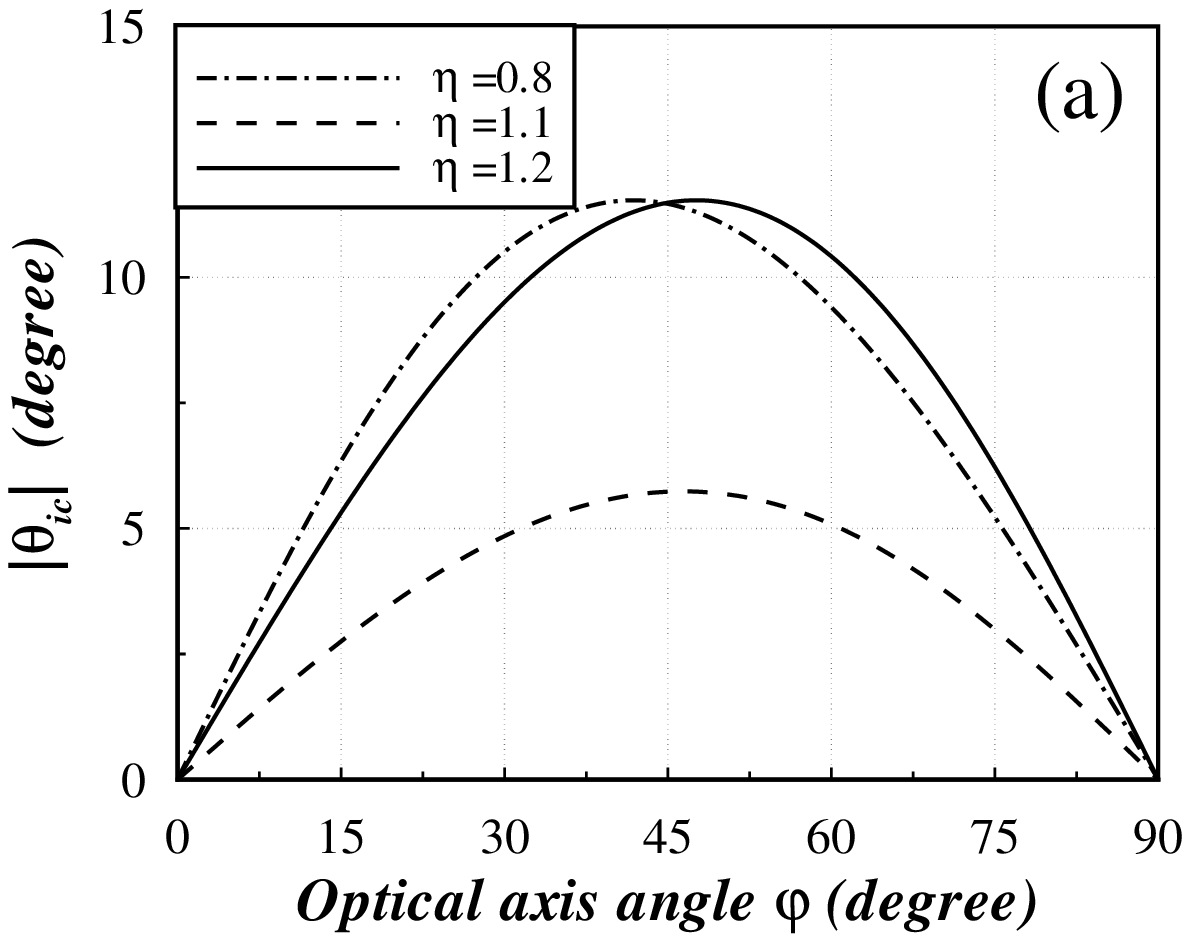}
\includegraphics[width=8cm]{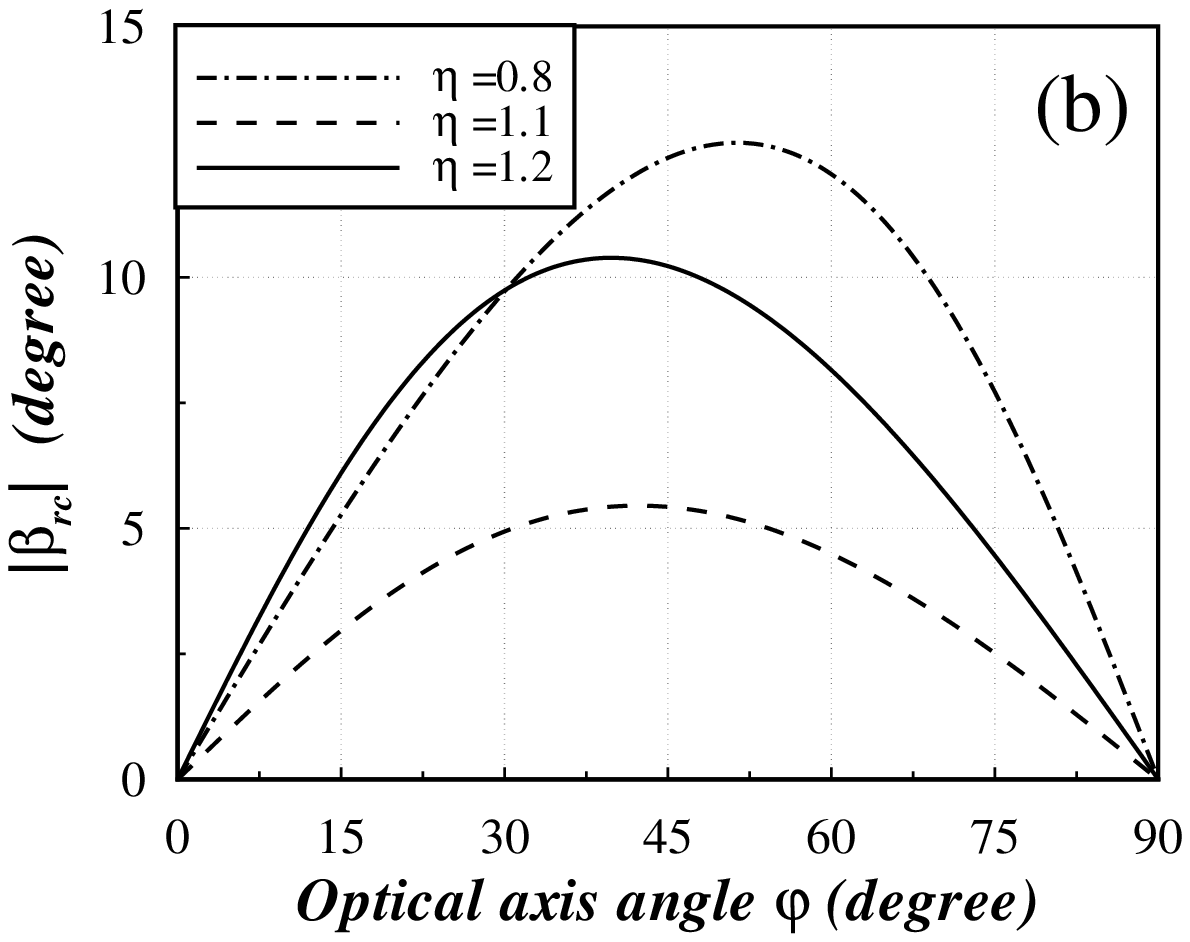}
\caption{\label{Fig3} (a) The dependence of the particular angle $
|\theta_{ic}|$ on the optical axis angle $\varphi$ for several
typical anisotropic media. (b) The dependence of the maximum
bending angle $ |\beta_{rc}|$ on the optical axis angle $\varphi$
for several typical anisotropic media, where
$\eta=\sqrt{\epsilon_\parallel/\epsilon_\perp}$ is the anisotropy
parameter.}
\end{figure}

It is noted that the negative refraction occurs in the fourth
quadrant for a $YVO_4$ crystal because it is a positive uniaxial
crystal, i.e. $\epsilon_\perp < \epsilon_\parallel$. Negative
refraction should occurs in the second quadrant for a negative
uniaxial crystal. However, The conditions for yielding anomalous
refractions, which refer to negative refraction in an anisotropic
PIM and positive refraction in an anisotropic NIM in following,
can be written uniformly as
\begin{equation}
0\leq |\theta_i| \leq |\theta_{ic}|, \label{region1}
\end{equation}
where $ \theta_{ic}$ is the particular  incident angle when the
ray refractive angle $\beta_r=0$. For both anisotropic PIM and
anisotropic NIM, $\theta_{ic}$ can be solved from
Eq.(\ref{energy}) under the condition $S_{tx}=0$, i.e.
\begin{equation}
\theta_{ic}=\sin^{-1}\left[\frac{\sigma(
\epsilon_\parallel-\epsilon_\perp)\mu\sin \varphi\cos\varphi
}{\sqrt{\epsilon_I
(\epsilon_\perp\mu\sin^2\varphi+\epsilon_\parallel\mu\cos^2\varphi)}}\right],\label{ic}
\end{equation}
The particular  incident angle $\theta_{ic}$ is dependent on the
optical axis angle $\varphi$, as well as the permittivities
$\epsilon_\perp$ and $\epsilon_\parallel$, and the permeability
$\mu$. Under the isotropic limit $\epsilon_\perp =
\epsilon_\parallel$ , one can obtain $\theta_{ic}=0$. It recovers
the well-known fact that no negative refraction can be observed at
the interfaces between two isotopic PIMs, and no positive
refraction between an isotropic PIM and an isotropic NIM. The
dependence of the particular  incident angle $| \theta_{ic}|$ on
the optical axis angle $\varphi$ for several typical uniaxial
crystal are shown in Fig.3(a). It can be seen when
$\varphi=0^\circ$ or $90^\circ$, the particular  incident angle $
\theta_{ic}$ approaches to zero, which means there are no
amphoteric refraction to be observed. For most anisotropic media,
when $\varphi$ is about $45^\circ$, $ |\theta_{ic}|$ has its
maximum value. To obtain the maximum of the particular  incident
angle $\theta_{ic}$, the optical axis angle $\varphi$ should be
chosen as
\begin{equation}
\varphi_{opt}=\cos^{-1}\left[\sqrt\frac{1}{1+\eta}\right].\label{i0opt}
\end{equation}
And the maximum incident angle $\theta_{ic}^{max}$ is
\begin{equation}
\theta_{ic}^{max}=
\sin^{-1}\left[\frac{\sigma\sqrt{\epsilon_\perp\mu}}{\sqrt{\epsilon_I}}(\eta-1)\right],
\label{icmax}
\end{equation}
where $\eta=\sqrt{\epsilon_\parallel/\epsilon_\perp}$ is the
anisotropy parameter. The variations of the optimal optical axis
angle $\varphi_{opt}$ and the maximum incident angle
$\theta_{ic}^{max}$ dependent on the anisotropy parameter $\eta$
are shown in Fig. 4, where we choose $\epsilon_I=\epsilon_\perp
\mu$. As an example by choosing $YVO_4$ crystal, we can obtain the
optimal optical axis angle of $\varphi_{opt}=46.57^\circ$ and the
corresponding maximum incident angle of
$\theta_{ic}^{max}=13.48^\circ$.

\begin{figure}
\includegraphics[width=10cm]{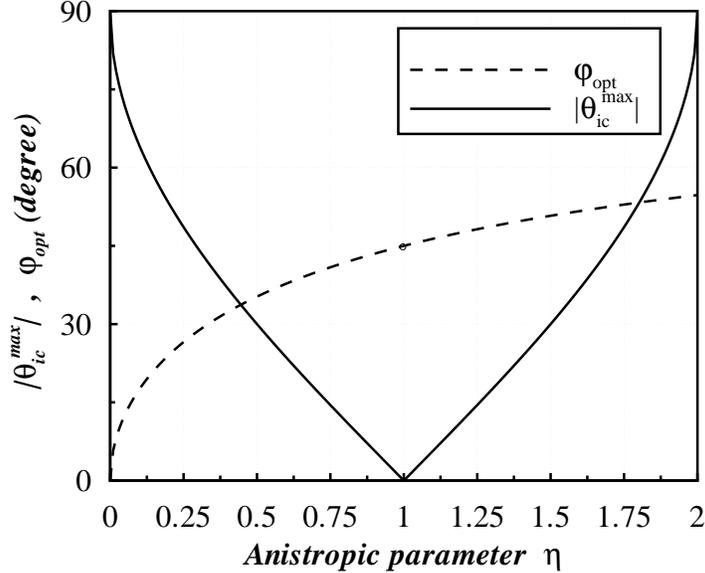}
\caption{\label{Fig4}The optimal optical axis angle
$\varphi_{opt}$ and the maximum incident angle
$|\theta_{ic}^{max}|$ as a function of the anisotropy parameter
$\eta$, where $\sqrt{\epsilon_I}=\sqrt{\epsilon_\perp\mu}$.}
\end{figure}

We can see that for a strongly anisotropy media, the maximum
incident angle could be very large. If there were an anisotropic
media satisfied with the condition $(
\sqrt{\epsilon_\parallel\mu}-\sqrt{\epsilon_\perp\mu}
)=\sqrt{\epsilon_I}$, the maximum incident angle could be
$|\theta_{ic}^{max}|=90^\circ$. When $(
\sqrt{\epsilon_\parallel\mu}-\sqrt{\epsilon_\perp\mu}
)>\sqrt{\epsilon_I}$, the Eq. (\ref{icmax}) is invalid and one
still has $|\theta_{ic}^{max}|=90^\circ$. In our knowledge, no
existing crystal can satisfy this condition. But we confide some
artificial anisotropy media could be constructed by using new
technique of liquid crystal, photonic crystal, etc, if it is worth
in practice.

\section{The strongest anomalous refraction}
It is also worth noting the particular  refractive angle
$\beta_{rc} $, which is the refraction angle of light when the
incident angle is zero or $k_{tx}=0$. It represents the strongest
negative refraction for anisotropic PIM or the strongest positive
refraction for anisotropic NIM. This angle is also the particular
incident angle of anomalous refraction when the light propagates
from an anisotropic media toward an isotropic media. The phenomena
of anomalous refraction at the interface from an isotropic NIM
toward an anisotropic PIM or PIM and its analysis are similar. So
does the anomalous refraction from an anisotropic media toward an
isotropic media.

The particular refractive angle $\beta_{rc} $ can be solved from
Eq.(\ref{energy}) under the condition $k_{tx}=0$, i.e.
\begin{equation}
\beta_{rc}=\tan^{-1}\left[\frac{ ( \eta^2-1) \sin
\varphi\cos\varphi}{\cos^2\varphi+\eta ^2\sin^2\varphi}\right].
\label{rc}
\end{equation}
It is also displayed as a function of the anisotropy parameter
$\eta$ and the optical axis angle $\varphi$. The dependence of the
particular  refractive angle $|\beta_{rc}|$ on the optical axis
angle $\varphi$ for several typical anisotropic media is shown in
Fig.3(b). The variations of particular refractive angle
$\beta_{rc}$ are similar to that of the particular  incident angle
$\theta_{ic}$. To obtain the maximum of the particular refractive
angle $\beta_{rc}$, the optical axis angle $\varphi$ should be
chosen as
\begin{equation}
\varphi_{opt}^\prime=\cos^{-1}\left[\sqrt\frac{1}{\eta^2+1}\right],
\label{t0opt}
\end{equation}
and the maximum bending angle $\beta_{rc}^{max}$ is
\begin{equation}
\beta_{rc}^{max}=\tan^{-1}\left[\frac{(\eta^2-1)}{2\eta}\right].
\label{rcmax}
\end{equation}

\begin{figure}
\includegraphics[width=10cm]{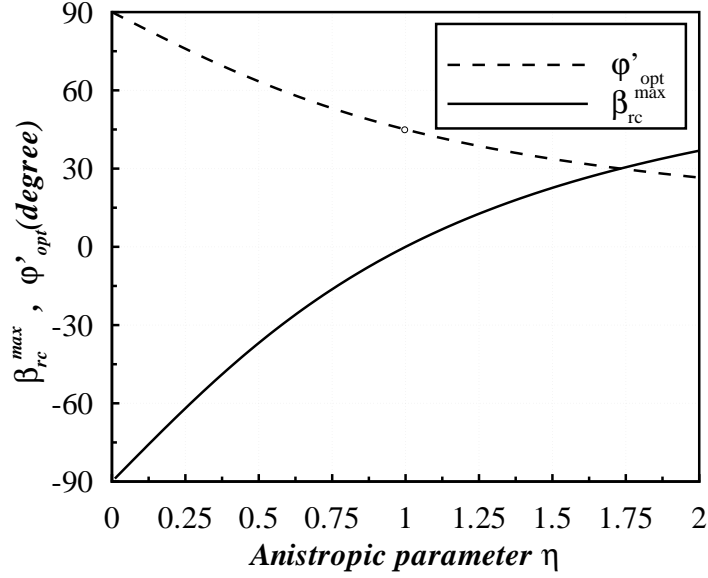}
\caption{\label{Fig5}The optimal optical axis angle
$\varphi_{opt}'$ and the maximum bending angle $\beta_{rc}^{max}$
as a function of the anisotropy parameter $\eta$.}
\end{figure}

Unlike the maximum incident angle $\theta_{ic}^{max}$, the max
bending angle $ \beta_{rc}^{max}$ is only dependent on anisotropy
parameter $\eta$ but independent on $\epsilon_I$. The variations
of the optimal optical axis angle $\varphi_{opt}^\prime$ and the
maximum bending  angle $\beta_{rc}^{max}$ dependent on the
anisotropy parameter $ \eta$  are shown in Fig.5.  Under the
isotropic limit $\eta=1$, $\beta_{rc}^{max}=0$, which recovers the
fact that no negative refraction can be observed at the interfaces
between two isotopic PIMs, and no positive refraction between an
isotropic PIM and an isotropic NIM.

\section{Discussions and conclusions}
Theoretical investigation above indicates that the amphoteric
refraction can be observed at a planar interface between an
isotropic media and an anisotropic media. When certain conditions
are satisfied, the interface between an isotropic PIM and an
anisotropic PIM (or NIM) can exhibit negative (or positive)
refraction. It is worth noting the anomalous refraction phenomenon
occurs only in a narrow range of incident angles for an
anisotropic PIM. Even for a very strongly anisotropy parameter,
i.e.
$(\sqrt{\epsilon_\parallel\mu}-\sqrt{\epsilon_\perp\mu})=\sqrt{\epsilon_I}$,
the incident angle is limited in the half planer, i.e.
$0^\circ<\theta_{i}<90^\circ$ or $0^\circ >\theta_{i} > 90^\circ$.

An intriguing phenomenon associated with the negative refraction
that has generated considerable interest is partial focusing by an
anisotropic NIM planar slab\cite{Smith2004a,Smith2004b}. With the
suitable arrangement of the optical axis and the appropriate
selection of anisotropy parameter of the anisotropic NIM, the
optimal focusing for anisotropic NIM as lens structures can be
effectively obtained. Obviously the anomalous positive refraction
may impair or weaken the focusing property. We do not expect that
the negative refraction in an anisotropic PIM may obtain the same
results. However, we believe that some potential applications can
be derived based on the amphoteric refractive properties in
anisotropic PIM or NIM. They can, for example, be used to provide
bending, angular dispersion, energy filtering, and beam
collimating for optical devices.

\begin{acknowledgements}
The authors are sincerely grateful to Prof. Qi Guo and L.B. Hu for
many fruitful discussions. This work is partially supported by
projects of the National Natural Science Foundation of China (No.
60278013), the Team Project of Natural Science Foundation of
Guangdong Province (No. 20003061), the Foundation of National
Hi-Tech Committee, the Fok Yin Tung High Education Foundation of
the Ministry of Education of China (No. 81058).
\end{acknowledgements}


\begin{references}

\bibitem{Veselago1968}    V.G. Veselago, Sov. Phys. Usp. {\bf 10},  509(1968).

\bibitem{Smith2000}  D. R. Smith ,  W. J. Padilla, D. C. Vier, S. C. Nemat-Nasser
and S. Schultz, \prl {\bf{84}}, 4184(2000).

\bibitem{Shelby2001} R. A. Shelby, D. R. Smith, and S. Schultz, Science {\bf 292},
77 (2001).

\bibitem{Pendry2000}   J. B. Pendry, \prl {\bf 85}, 3966 (2000).

\bibitem{Valanju2002}P. M. Valanju, R. M. Walser, and A. P. Valanju \prl {\bf 88},
187401 (2002).

\bibitem{Hooft2001}G.W. Hooft, \prl {\bf 87}, 249701(2001); J. B. Pendry,
\prl {\bf 87} 249702(2001);

\bibitem{William2001}J. M. Williams, \prl {\bf 87}, 249703(2001); J. B. Pendry,
\prl {\bf 87}249704(2001)

\bibitem{Garcia2002} N. Garcia and M. Nieto-Vesperinas \prl {\bf 88} 207403(2002)

\bibitem{Parazzoli2003}  C.G. Parazzoli, R. G.Greegpr, K. Li, B. E. C. Koltenba,
 and M. Tanielian,\prl {\bf 90}, 107401 (2003).

\bibitem{Smith2002}   D. R. Smith, D. Schurig and J. B. Pendry, \apl {\bf 81}, 2713 (2002).

\bibitem{Pacheco2002}  J. Pacheco, Jr, T. M. Grzegorczyk,  B. I. Wu, Y. Zhang and
J. A. Kong, \prl {\bf 89}, 257401 (2002).


\bibitem{Foteinopoulou2003}S. Foteinopoulou, E. N. Economou, and C.M. Soukoulis,
\prl {\bf 90}, 107402 (2003)

\bibitem{Houck2003} A. A. Houck, J. B. Brock, and I. L. Chuang, \prl {\bf 90}, 137401
(2003)

\bibitem{Santos2003}G. Gomez-Santos, \prl {\bf 90}, 077401(2003).

\bibitem{Feise2002}Michael W. Feise, Peter J. Bevelacqua, and John B. Schneider \prb
{\bf  66}, 035113(2002)

\bibitem{Notomi2000} M. Notomi, \prb {\bf 62}, 10696 (2000).


\bibitem{Luo2002} C. Luo, S. G. Johnson, J. D.  Joannopoulos, and J. P. Pendry,
\prb {\bf 65}, 201104 (2002).


\bibitem{Lindel2001}   I. V. Lindel, S. A. Tretyakov, K. I. Nikoskinen, and S. Ilvonen,
Opt. Technol. Lett. {\bf 31}, 129(2001).

\bibitem{Zhang2003}   Y. Zhang, B. Fluegel, and A. Mascarenhas, \prl {\bf 91}, 157401(2003).

\bibitem{Liu2004}  Z. Liu, Z. Lin, and S. T. Chui, \prb {\bf 69}, 115402 (2004).

\bibitem{Hu2002}   L. B. Hu and S. T. Chui, \prb {\bf 66}, 085108 (2002).

\bibitem{Smith2003}   D. R. Smith and D. Schurig, \prl {\bf 90}, 077405 (2003).

\bibitem{Yariv1984}    A. Yariv and P. Yeh, Optical Wave in Crystals,
(John Wiely  Sons,New York, 1984) Chap.VI.

\bibitem{Smith2004a}  D. R. Smith, D. Schurig, J. J. Mock, P. Kolinkp and  P. Rye
, \apl {\bf 84},  2244 (2004).

\bibitem{Smith2004b}  D. R. Smith,P. Kolinkp and D. Schurig, \josab {\bf 21},  1032
(2004).

\end{references}
\end{document}